\documentclass[fleqn,usenatbib]{mnras}

\usepackage[T1]{fontenc}

\DeclareRobustCommand{\VAN}[3]{#2}
\let\VANthebibliography\thebibliography
\def\thebibliography{\DeclareRobustCommand{\VAN}[3]{##3}\VANthebibliography}

\usepackage{graphicx}	
\usepackage{amsmath}	
\usepackage{amssymb}	

\title[A helium nova in the LMC]{A helium nova in the Large Magellanic Cloud
--- the faint supersoft X-ray source [HP99]159}

\author[Kato, Hachisu, \& Saio]{
Mariko Kato,$^{1}$\thanks{E-mail:mariko.kato@hc.st.keio.ac.jp}
Izumi Hachisu,$^{2}$
and Hideyuki Saio$^{3}$
\\
$^{1}$Department of Astronomy, Keio University,
Hiyoshi, Kouhoku-ku, Yokohama 223-8521, Japan\\
$^{2}$Department of Earth Science and Astronomy,
College of Arts and Sciences, The University of Tokyo, 3-8-1 Komaba,
Meguro-ku, Tokyo 153-8902, Japan\\
$^{3}$Astronomical Institute, Graduate School of Science,
    Tohoku University, Sendai 980-8578, Japan
}

\date{Accepted XXX. Received YYY; in original form ZZZ}

\pubyear{2023}

\begin{document}
\label{firstpage}
\pagerange{\pageref{firstpage}--\pageref{lastpage}}
\maketitle

\begin{abstract}
We propose a helium nova model for the Large Magellanic Cloud (LMC) supersoft 
X-ray source (SSS) [HP99]159.  This object has long been detected as a faint 
and persistent SSS for about 30 years, and recently been interpreted to be
a source of steady helium-shell burning, because no hydrogen lines
are observed.  
We find that the object can also be interpreted as in 
a decaying phase of a helium nova.
The helium nova is slowly decaying toward the quiescent phase, 
during which the observed temperature, luminosity, and SSS lifetime
($\gtrsim 30$ years) are consistent with a massive white dwarf model of
$\sim$ 1.2 $M_\odot$.  If it is the case,  
this is the second discovery of a helium nova outburst
after V445 Pup in our Galaxy and also the first identified helium nova
in the LMC.  We also discuss the nature of the companion
helium star in relation to Type Ia supernova progenitors.
\end{abstract}




\section{Introduction}


[HP99]159 \citep{hp99} is an LMC X-ray source that has been observed
since the 1990.  \citet{gre23} reported detailed observational properties
and characterized this object as a binary consisting of an X-ray emitting
white dwarf (WD) and a hydrogen deficient companion star.

The X-ray spectrum taken on April 1992 with
ROSAT shows a blackbody temperature of k$T =38 \pm 15$ eV
and unabsorbed bolometric luminosity of
$L_X= 1.3^{+41.7}_{-1.0} \times 10^{36}$ erg s$^{-1}$.
XMM-NEWTON observed [HP99]159 on 16/17 September 2019 and its
spectrum yields k$T =45 \pm 3$ eV
and $L_X= 6.8^{+7.0}_{-3.5} \times 10^{36}$ erg s$^{-1}$
for the distance of LMC (50 kpc).
eROSITA scanned the region including [HP99]159 five times,
and the spectrum fits show k$T=$ 42 -- 44 eV.

These temperatures suggest that the X-ray emitting source is a hot WD.
\citet{gre23} estimated the WD mass to be $M_{\rm WD}= 
1.2^{+0.18}_{-0.4} M_\odot$ from a mass versus radius relation of WDs.
\citet{gre23} also obtained optical spectra in August -- October 2020
that show no indication of hydrogen lines, suggesting a helium star companion. 
Moreover, the spectra show no broad emission lines,
that means the absence of strong mass loss.
The orbital period was determined to be $P_{\rm orb}=2.33$ day
(or 1.16 day).

From these observational properties, 
\citet{gre23} concluded that the X-ray source [HP99]159 is
a steady helium-burning WD accreting from a helium donor star.

Such a helium-accreting massive WD is one of the progenitor
systems of Type Ia supernovae 
\citep{ibe94,wan09mc,mcc14jf,gui10,hil16,wan17,kat18hvf}. 
\citet{koo23} recently reported the discovery of such a Type Ia supernova,
SN2020eyj.  Their spectra clearly show helium-rich, hydrogen-deficient
circumstellar material, so that this SN is the first definite Type Ia 
supernova whose progenitor is a binary consisting of a WD and a helium
star donor.
Together with the discovery of the hydrogen-deficient X-ray source [HP99]159,
we could enlarge the possibility of helium-donor channel toward 
a Type Ia supernova. 

\citet{gre23} interpreted that the X-ray luminosity comes from
steady helium-shell-burning on a WD. However, 
the X-ray flux is too faint to be
compatible with that of steady helium-shell-burning, about 100 times smaller.
The observed X-ray luminosity $\sim 1800 ~L_\odot$ is much lower
than that of steady helium-shell-burning ($\gtrsim 20,000~L_\odot$). 
\citet{gre23} considered this faint X-ray flux as
the steady helium-burning with a mass-accretion rate of
$\dot M_{\rm acc}=1.5 \times 10^{-7}~M_\odot$ yr$^{-1}$.
However, with this small mass-accretion rate, helium burning
is unstable and results in repeated helium nova outbursts
\citep{kat08v445pup,kat18hvf}.

\begin{figure}
	\includegraphics[width=\columnwidth]{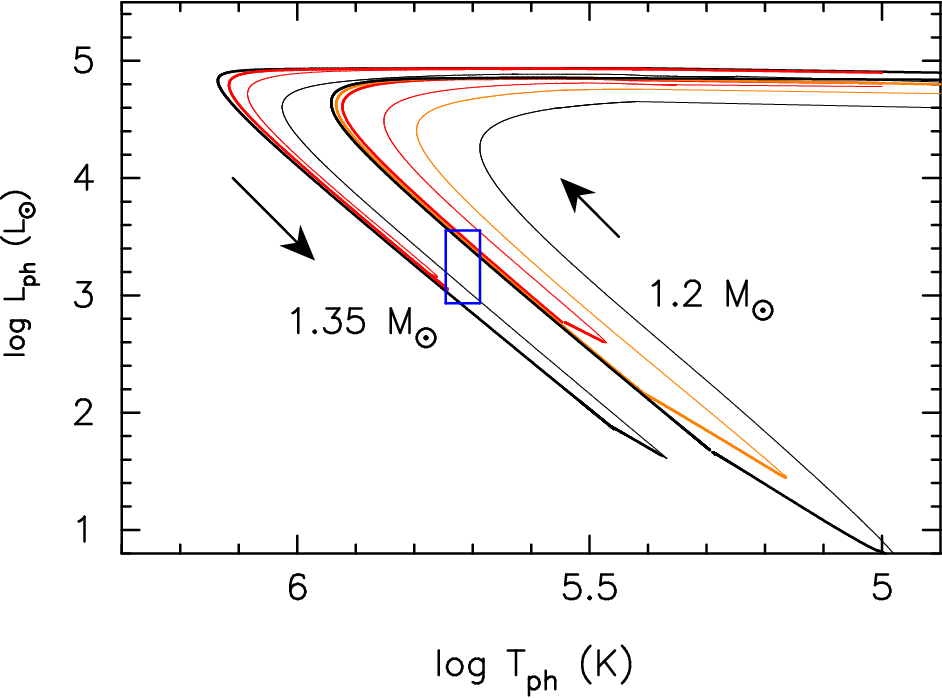}
    \caption{
The HR diagram for He shell flashes on a 
1.35 $M_\odot$ WD with mass-accretion rates of helium
${\dot M}_{\rm acc}= 1.6\times 10^{-7}~M_\odot$ yr$^{-1}$ (black line)
or $7.5\times 10^{-7}~M_\odot$ yr$^{-1}$ (red line),
and a 1.2 $M_\odot$ WD with
$1.6\times 10^{-7}~M_\odot$ yr$^{-1}$ (black line),
$3.0\times 10^{-7}~M_\odot$ yr$^{-1}$ (orange line), or
$6.0\times 10^{-7}~M_\odot$ yr$^{-1}$ (red line).
The thick lines represent the decay phases whereas the
thin lines the rising phases. The blue rectangular denotes the observed 
ranges of the luminosity and temperature of [HP99]159 \citep{gre23}.}
    \label{fig:hrcompa}
\end{figure}

In this letter, we propose the alternative to their interpretation,
the decay phase of a helium nova.
Helium novae were theoretically predicted by \citet{kat89} as a
nova outburst caused by a helium shell flash on a WD.
It had long been a theoretical object until the discovery
of the helium nova V445 Pup 2000 in our Galaxy \citep{kan00,ash03}.
Since then, no further helium novae have been identified yet.

Theoretically, a helium nova evolves similarly to
a classical nova: it brightens up and reaches
optical maximum, the optical magnitude gradually decays,
followed by the supersoft X-ray source (SSS) phase.
In V445 Pup, a strong dust blackout occurred
200 days after the optical peak, so we could not observe
the SSS phase of this helium nova.
If [HP99]159 is a helium nova, it gives us invaluable information on the
late phase of a helium nova outburst.
Furthermore, [HP99]159 is the first identified helium nova in the LMC.

\begin{figure}
	\includegraphics[width=\columnwidth]{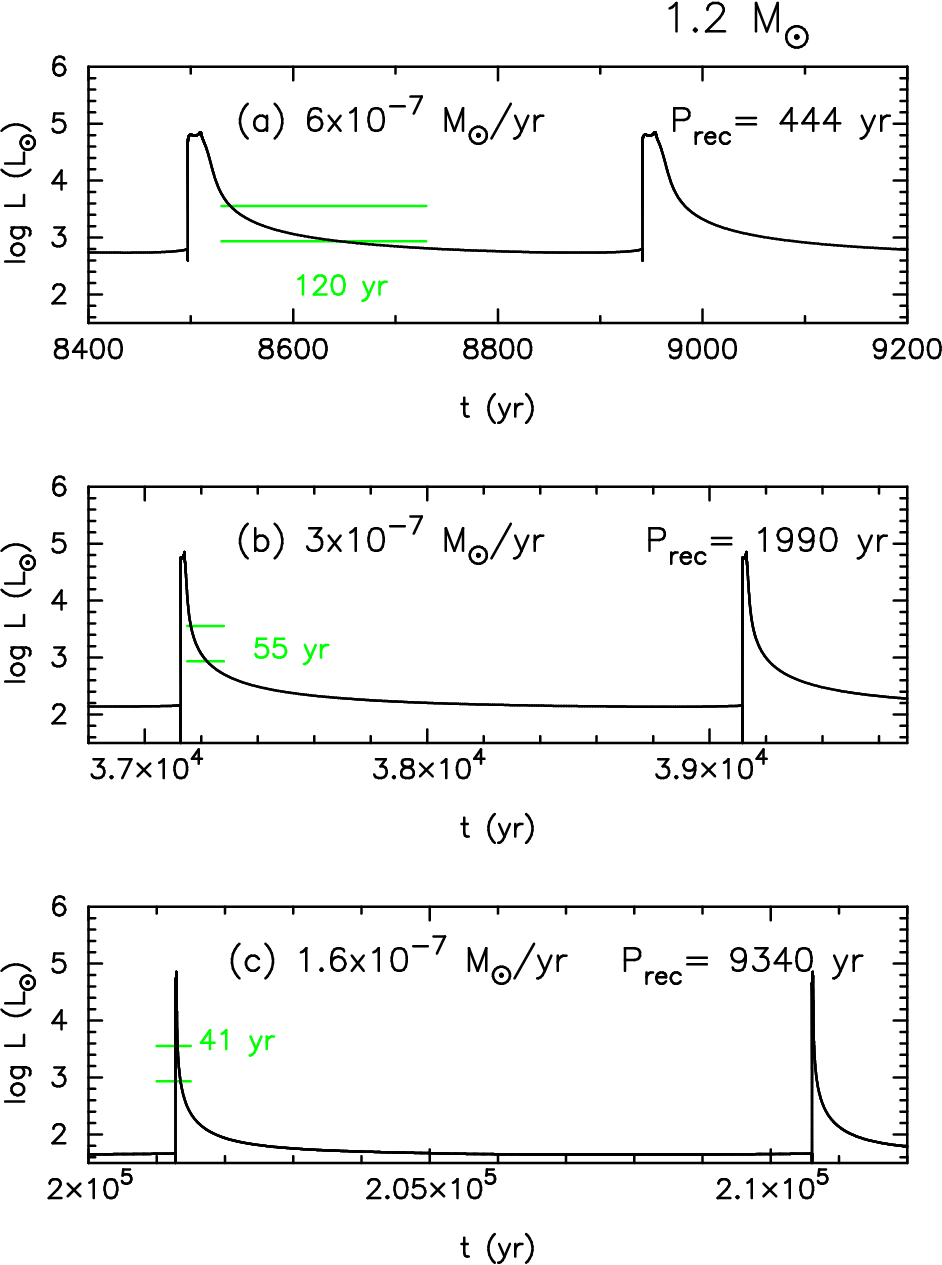}
    \caption{
Temporal variation of the photospheric luminosity of helium shell flashes
for our 1.2 $M_\odot$ WD models with the mass-accretion rates of helium
(a) $\dot {M}_{\rm acc}= 6.0\times 10^{-7}$,
(b) $3 \times 10^{-7}$, and
(c) $1.6\times 10^{-7}~M_\odot$ yr$^{-1}$.
Two green horizontal lines indicate the upper and lower values of the 
error box in Figure \ref{fig:hrcompa},   
which is obtained with XMM-Newton in 2019. 
The upper limit obtained with ROSAT in 1992  
is much higher, i.e., $\log L(L_\odot) \le 4.0$
(see Fig. \ref{fig:hp99x}). 
 \label{fig:Lcomparim12}}
\end{figure}


\begin{figure}
        \includegraphics[width=\columnwidth]{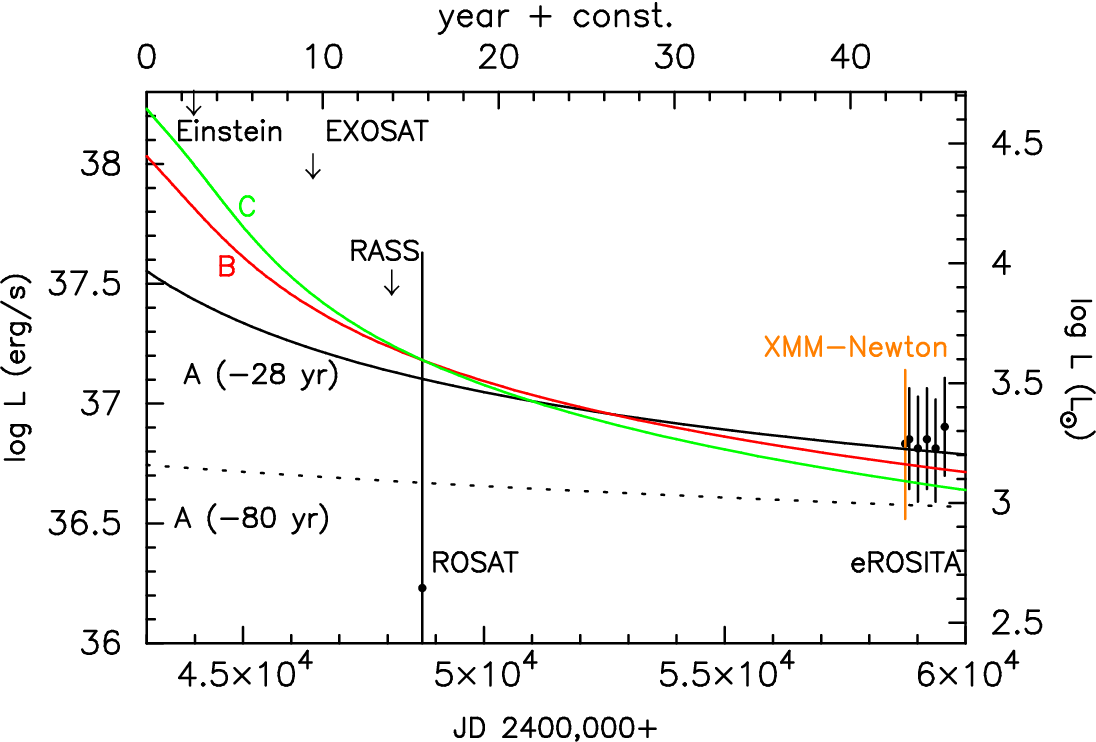}
    \caption{Comparison of the three light curves of the 1.2 $M_\odot$ WD 
in Figure \ref{fig:Lcomparim12} with observation \citep{gre23}. 
The three downward arrows indicate 
the upper limits obtained with Einstein, EXOSAT, and 
RASS (ROSAT all sky survey). The data labeled ROSAT is obtianed 
with ROSAT-PSPC (pointed observation). 
The two black lines labeled A correspond to the same model 
in Fig. \ref{fig:Lcomparim12}(a) 
but went into outburst 28 years ago (solid line), or 80 years ago 
(dotted line) from $t=0$ yr on the upper abscissa, i.e., at $t=-28$ yr,
or $t=-80$ yr, respectively. 
The red line indicats the model in Fig. \ref{fig:Lcomparim12}(b) 
that went into outburst at $t=-19$ yr, and 
the green line indicates the model 
in Fig. \ref{fig:Lcomparim12}(c) at $t=-14$ yr.} 
\label{fig:hp99x}
\end{figure}

\section{Model Light Curves}


We apply the helium flash models and He star evolution models
already published in \citet{kat18hvf} and \citet{kat08v445pup}, 
respectively, to the [HP99]159 X-ray source.
In these model calculations, we assumed spherical symmetry and
used a Henyey-type evolution code. 
For helium shell flashes, if we start
our calculation from an arbitrary initial condition, 
we need time-consuming calculations for a huge number of shell flashes
until the shell flash properties approach a limit cycle.
To avoid such a lengthy task, we adopt the initial WD models that are 
in a thermal equilibrium with the assumed mass-accretion rate. 
Then, we need only several shell flashes to reach almost a limit cycle.
A typical mesh number is about 2,000. Nucleosynthesis in the helium burning
is calculated up to $^{28}$Si. 
When the nova envelope expands to a giant size, we assume 
a mass-loss from the helium-rich envelope
to avoid numerical difficulties \citep{kat17}.
We use the OPAL opacity tables \citep{igl96}.

The chemical composition of accreting matter to the WD is assumed to be
$X=0.0$, $Y=0.98$, and $Z=0.02$, although
[HP99]159 is located in the LMC, a less metal-enriched galaxy 
\citep[e.g.,][for an age-metallicity relation of the LMC]{pia13g}.
A smaller $Z$ may result in a somewhat larger ignition mass that 
strengthens thermonuclear runaway and wind mass-loss.  But, this affects
the He nova evolution only in a very early phase.  After that,
the helium-rich nova envelope approaches steady-state, in which
the nuclear energy release rate is balanced with the 
radiative loss and gravitational energy release. 
As a result, the smaller $Z$ hardly affects the evolution 
because the nuclear burning rate of $3\alpha$ does not depend on the $Z$.

Figure \ref{fig:hrcompa} shows one cycle of shell flashes in the HR diagram
for 1.35 $M_\odot$ and 1.2 $M_\odot$ WDs with different mass-accretion rates.
For a smaller mass-accretion rate, the locus of one cycle goes outside
especially in the rising phase (thin line parts).
The blue error box indicates the range of bolometric luminosity
$L_{\rm X}= 6.8^{+7.0}_{-3.5} \times 10^{36}$ erg s$^{-1}$ and
temperature k$T =45 \pm 3$ eV obtained by \citet{gre23}
for [HP99]159.
The position of the error box is consistent with the decay phase of
both the 1.35 $M_\odot$ and 1.2 $M_\odot$ WDs.
More massive ($> 1.35 M_\odot$) or less massive ($ < 1.2 M_\odot$) WDs
are excluded by this constraint.

Figure \ref{fig:Lcomparim12} shows three theoretical light curves
of helium novae for the 1.2 $M_\odot$ WD
with three mass-accretion rates.
An optically bright phase of a nova outburst corresponds to
the first half of the high luminosity phase ($L \sim 10^5 ~L_\odot$
and $\log T_{\rm ph} \la 5.5$) in Figure \ref{fig:hrcompa},
while a low luminosity decay phase in Figure \ref{fig:hrcompa} 
is related to a long lasted low luminosity period 
($L \la 10^3 ~L_\odot$) in Figure \ref{fig:Lcomparim12}.
The SSS phase in Figure \ref{fig:Lcomparim12}
begins in the later half of the high luminosity phase 
($L \sim 10^5 ~L_\odot$ and $\log T_{\rm ph} \ga 5.5$)
in Figure \ref{fig:hrcompa}, and continues until 
the luminosity substantially decreases.
The recurrence period is longer for a smaller mass-accretion rate.
The photospheric luminosity reaches $L_{\rm ph} \sim 10^5 ~L_\odot$ at
the flash peak and gradually decreases after that.
We indicate the upper and lower limits for
the bolometric luminosity of [HP99]159,
$L_{\rm ph}\approx L_{\rm X}= 6.8^{+7.0}_{-3.5} \times 10^{36}$ erg s$^{-1}$
\citep{gre23}, and the theoretical duration in the above range of $L_{\rm X}$.
All of the three models satisfy the SSS duration of 
$\gtrsim \tau_{\rm SSS} \sim 30$ yr.
Here, $\tau_{\rm SSS}$ is the lifetime of [HP99]159 as a very faint
SSS. [HP99]159 has been observed since
the first positive detection with ROSAT in 1992.
 
Figure \ref{fig:hp99x} shows the observed X-ray fluxes summarized
by \citet{gre23}. 
With the three upper limits from Einstein, EXOSAT, and ROSAT (labeled RASS),
we assume that [HP99]159 has kept almost constant luminosity during the 
last 40 years. 
The three $1.2~M_\odot$ WD models in Figure \ref{fig:Lcomparim12}
are consistent with the long term SSS observation.
We plot these three models in Figure \ref{fig:hp99x}. 
The two models of $\dot M_{\rm acc}=1.6\times 10^{-7}~M_\odot$ yr$^{-1}$
and $3\times 10^{-7}~M_\odot$ yr$^{-1}$ show 
short duration of $\tau_{\rm SSS}= 41$ years and 55 years, 
respectively, that is barely consistent with observed range. 
The model of $\dot M_{\rm acc}=6\times 10^{-7}~M_\odot$ yr$^{-1}$ shows 
a slow decay of $\tau_{\rm SSS}= 120$ years. 
The solid black line in Figure \ref{fig:hp99x} shows 
its early decay phase, while the black dotted line is for a late decay phase,
52 years later than the solid black line. 
Thus, our 1.2 $M_\odot$ WD with $\dot M_{\rm acc}=6\times 10^{-7}~M_\odot$
yr$^{-1}$ naturally explains all of the X-ray properties of [HP99]159
summarized by \citet{gre23}. 
 
We searched archives for a corresponding optical outburst 
but found no information on [HP99]159 in ADS, ATel, and AAVSO. 
The search for an otpical counter part in individual old plates is 
far beyond the scope of this work. 

We have examined the 1.35 $M_\odot$ WD models with three different
mass-accretion rates of ${\dot M}_{\rm acc}= 7.5\times 10^{-7}$,
$3 \times 10^{-7}$, and $1.6\times 10^{-7}~M_\odot$ yr$^{-1}$
in Figure \ref{fig:hrcompa} \citep{kat18hvf}.
All of the models show $\tau_{\rm SSS} \lesssim 30$ yr 
and could not satisfy the observational constraints in Figure \ref{fig:hp99x}.

\begin{figure}
        \includegraphics[width=\columnwidth]{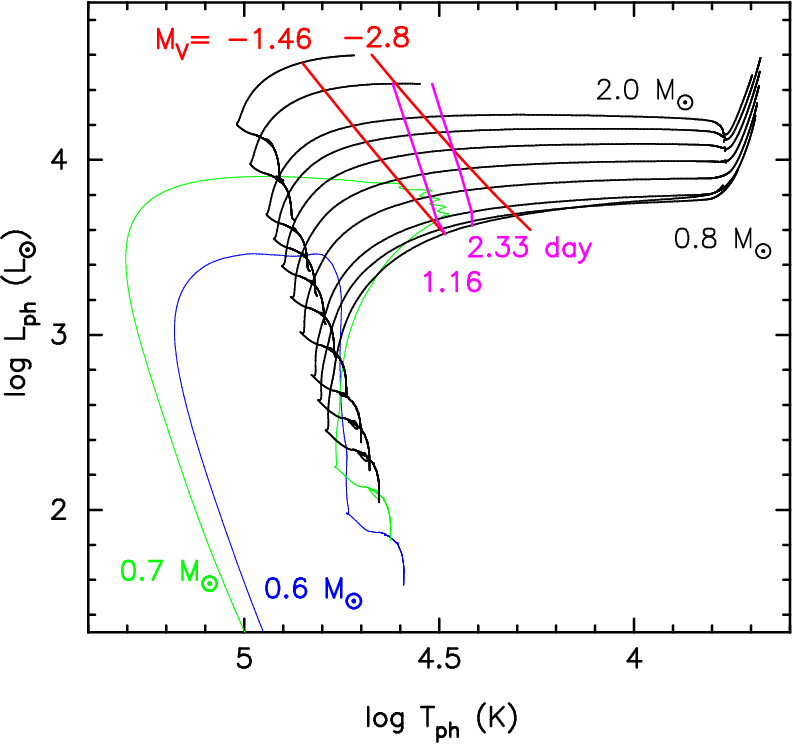}
    \caption{HR diagram for various mass He star evolutions.
Tracks are plotted by the black lines
for the mass, from top to bottom,
3.0, 2.5, 2.0, 1.8, 1.6, 1.4, 1.2, 1.0, 0.9, and 0.8 $M_\odot$.
The 0.7 and 0.6 $M_\odot$ are denoted by the green and blue lines,
respectively, and do not evolve to the red giant regime.
All of the data are taken from \citet{kat08v445pup}.
The red lines indicate the positions of $M_V=-1.46$ and $-2.8$.
The pink lines connect each helium star model that 
just fills its Roche-lobe for a binary of a 1.2 $M_\odot$ WD 
and He star with $P_{\rm orb}=1.16$ and 2.33 day.}
    \label{fig:Hestarhr}
\end{figure}

\section{Discussion}

\citet{gre23} obtained the absolute $V$ magnitude of [HP99]159
to be $M_V=-2.8$ for the LMC distance (50 kpc).
They interpreted that an accretion disk dominates the $V$ brightness. 
The $V$ magnitude from the non-irradiated disk is, however, 
estimated to be as faint as $M_V=1.14$ for a 1.2 $M_\odot$ WD with 
$\dot M_{\rm acc} =1\times 10^{-6}~M_\odot$ yr$^{-1}$  
\citep[Equation (A5) in ][]{web87}. 
Here we assume that the binary is close to face-on 
\citep{gre23}.  

\citet{gre23} reported $\sim 0.2$ mag fluctuation in the 
$V$ and $I$ long-term light-curves of [HP99]159. 
The origin of the variation was not suggested in their paper,
but this $\sim 0.2$ mag variation reminds us the flickering, that are
often observed in disk-dominated cataclysmic-variables 
\citep{zam18,bru21}. If the variation in [HP99]159 is caused by
the flickering in the accretion disk, we may expect substantial 
contribution from the accretion disk in the optical band.

\citet{pop96} calculated optical spectra of 
an accretion disk and companion star both irradiated by a hot WD. 
Their composite spectra show an excess toward longer wavelength 
owing to such irradiation effects. 
A similar excess is also seen in the spectra of [HP99]159 \citep{gre23}. 
Thus, we regard that both the irradiated disk and companion star contribute 
to the V magnitude of [HP99]159. 

Figure \ref{fig:Hestarhr} shows evolutions of helium stars in the
HR diagram for various zero-age masses, 
taken from \citet{kat08v445pup}.
The low mass He stars of 0.6 and 0.7
$M_\odot$ do not evolve toward a helium red giant, but return to a
higher temperature region than that at zero-age,
whereas more massive stars evolve toward a red giant.
Note that the stellar mass is assumed to be constant, 
i.e., no mass loss is assumed.
Also irradiation effects are not included. 

This figure also shows the line of $M_V=-2.8$ 
calculated from $T_{\rm ph}$ and $L_{\rm ph}$
with a canonical response function of the $V$-band filter.
The line of $M_V=-1.46$ indicates a case if there are some 
contributions from the irradiation effects on the helium star 
and accretion disk as discussed below.
We also added a line of the orbital period $P_{\rm orb}=1.16$ day 
and 2.33 day, assuming a binary consisting of 
a 1.2 $M_\odot$ WD and a Roche lobe-filling helium star.
The crossing points of these two orbital period lines 
with the $M_V=-2.8$ line show the helium companion mass is 
2.5 $M_\odot$ and 1.6 $M_\odot$, respectively. 

These companion masses, however, seem to be too large.   
\citet{kat08v445pup} calculated the mass loss rate from a Roche 
lobe-filling helium star, assuming a constant lobe radius of 1.5 $R_\odot$
(see their Figure 8). 
Although the binary parameter is slightly different,
their results suggest that the mass-transfer rate from 
a Roche lobe-filling $ > 0.8~M_\odot$ helium star is as large as
$|\dot M| > 10^{-6}~M_\odot$ yr$^{-1}$. 
Helium burning is stable for such a high mass-accretion rate 
\citep[Figure 1 in ][]{kat18hvf}. 
A WD of steady helium burning is 
too bright (several $\times 10^4~L_\odot$), incompatible 
with the X-ray luminosity of [HP99]159 as in Figure \ref{fig:hrcompa}. 

\citet{kat08v445pup} showed that the mass transfer rate 
decreases from $> 10^{-6}~M_\odot$ yr$^{-1}$ 
finally to $10^{-7}~M_\odot$ yr$^{-1}$ 
when the mass of the donor helium star approaches 0.8 $M_\odot$. 
A 0.8 $M_\odot$ donor star has the brightness $M_V=-1.46$ for 
$P_{\rm orb}=1.16$ day and  $M_V=-2.03$ for $P_{\rm orb}=2.33$ day 
in Figure \ref{fig:Hestarhr}. 
The difference from $M_V=-2.8$ can be attributed to the 
irradiation effects on the helium star and accretion disk. 



\section{Conclusions}


We propose a helium nova model that satisfies observational aspects of 
[HP99]159: it is a binary consisting of a helium-accreting 
$\sim 1.2 ~M_\odot$ WD and a Roche lobe-filling, evolved helium star of 
$\sim 0.8-0.9 ~M_\odot$.
The X-ray flux comes from the photosphere of the still hot WD.
It is now cooling toward the quiescent phase after a helium nova outburst.
The mass-transfer rate onto the WD is a few to several 
$\times 10^{-7}~M_\odot$ yr$^{-1}$.  
The optical brightness $M_V=-2.8$ is the contribution not only from
the (irradiated) companion star but also from the irradiated disk.

We may conclude that [HP99]159 is the second identified helium nova
after V445 Pup and a key object 
in Type Ia supernova progenitor scenarios.

\section*{Acknowledgements}
 
We are grateful to the anonymous referee for useful comments, 
which improved the manuscript.


\section*{Data Availability}
The data underlying this article will be shared on reasonable request 
to the authors. 
 




\bibliographystyle{mnras}






\bsp	
\label{lastpage}
\end{document}